\documentclass
{mn2e}
\usepackage{epsfig}
\usepackage{amsmath, amssymb,bm}

\def \msun{\rm M_{\odot}}

\def \et{et al.} 
\begin{document}
\title[The End of the Black Hole Dark Ages, and Warm Absorbers]{The End of the
  Black Hole Dark Ages -- and the Origin of Warm Absorbers}

\author[A.~R.~King \& K.A. Pounds] 
{
\parbox{5in}{A.R.King$^{1, 2}$ \& K.A.Pounds$^1$}
\vspace{0.1in} 
 \\ $^1$ Department of Physics \& Astronomy, University of Leicester,
 Leicester LE1 7RH UK 
 \\ $^2$  Astronomical Institute `Anton Pannekoek', University of Amsterdam, Postbus 94249 NL--1090 GE Amsterdam, The Netherlands
}

\maketitle

\begin{abstract}
We consider how the radiation pressure of an accreting supermassive
hole (SMBH) affects the interstellar medium around it. Much of the gas
originally surrounding the hole is swept into a shell with a
characteristic radius somewhat larger than the black hole's radius of
influence ($\sim$ 1-100~pc). The shell has a mass directly comparable
to the ($M - \sigma$) mass the hole will eventually reach, and may
have a complex topology. We suggest that outflows from the central
supermassive black holes are halted by collisions with the shell, and
that this is the origin of the warm absorber components frequently
seen in AGN spectra.  The shell may absorb and reradiate some of the
black hole accretion luminosity at long wavelengths, implying both
that the bolometric luminosities of some known AGN may have been
underestimated, and that some accreting SMBH may have escaped
detection entirely.

\end{abstract}


\footnotetext[1]{E-mail: ark@astro.le.ac.uk}

\section{Introduction}
\label{intro}

Astronomers now generally agree that the centre of most galaxies
contains a supermassive black hole (SMBH). Active galactic nuclei
(AGN) correspond to phases when the hole is growing its mass by
accreting gas from a very small--scale disc around it. But is it not
immediately obvious why these phases are in practice directly
observable, as there are good reasons to expect that a significant
mass of gas largely surrounds the hole at such epochs. First, a
majority of AGN show significant signs of obscuration (cf the
discussion in Elvis, 2000). A large gas mass is needed close enough to
the hole to grow it in a reasonable time, and this must cover a
significant solid angle since the total mass of a geometrically thin
disc is severely limited by self--gravity. Finally, simple estimates
of the column densities of matter subject to a galactic potential
confirm the impression that most AGN are likely to be at least formed
in fairly dense gas environments.

This suggests that when we do see black holes accreting, they may have
perturbed the gravitational equilibrium of the matter which would
otherwise block our view. An obvious way of doing this is to push it
away, spreading it over a larger area and reducing its column
density. We investigate this idea here.

\section{Pushing for transparency}

In luminous AGN
far the strongest energy supply potentially pushing matter away
from a black hole is its accretion luminosity $L$. (This is of course
not true for low--luminosity radio galaxies, where jets may interact
with the surroundings.) By contrast,
accretion disc {\it winds} are generally limited to mechanical
luminosities $\eta L/2\simeq 0.05L$, where $\eta \sim 0.1$ is the
accretion efficiency (e.g. King \& Pounds, 2003; King, 2003), and much
of this is likely to be lost in shocks (King 2003, 2010).

If the surroundings have high optical depth to scattering (i.e. are
strongly obscuring), photons scatter many times, and so radiation
pressure must become significant. In scattering slightly
inelastically, the luminosity $L$ does work pushing against the
gravitational force on the surrounding gas in the central spheroid of
the galaxy. If the gas is not is large--scale dynamical motion,
we assume that it is distributed isothermally, i.e. with
density
\begin{equation}
\rho(r) = {f_g\sigma^2\over 2\pi Gr^2},
\label{rho}
\end{equation}
so that the gas mass within radius $R$ is
\begin{equation}
M_g(R) = {2f_g\sigma^2R\over G},
\label{gmass}
\end{equation}
and the total mass (including stars, and any dark matter) is 
\begin{equation}
M(R) = {2\sigma^2R\over G}.
\label{mass}
\end{equation}
Here $\sigma$ is the velocity dispersion and $f_g$ is the gas fraction
relative to all matter (e.g. dark matter, and stars)
which has cosmic value 0.16. We assume that $f_g$ does not
vary strongly across the central region of the galaxy.

The pressure of trapped radiation sweeps the gas up progressively into
a shell of inner radius $R$ and mass $M_g(R)$. If the shell is
geometrically thin its electron scattering optical depth at radius $R$ is
\begin{equation} 
\tau_{\rm sh}(R) = {\kappa M_g(R)\over 4\pi R^2} = {\kappa
  f_g\sigma^2\over 2\pi GR},
\end{equation}
where $\kappa \simeq 0.34~{\rm cm^2\, g^{-1}}$ is the electron scattering
opacity.
If the shell is geometrically thicker, $\tau_{\rm sh}(R)$ is an upper
limit to its optical depth, as on average the gas is more spread out
(i.e. at larger radii). The undisturbed gas outside $R$ has
optical depth
\begin{equation}
\tau(R) = \int_R^{\infty} \kappa \rho(r) {\rm d}r = {\kappa
  f_g\sigma^2\over 2\pi GR} = \tau_{\rm sh}(R),
\end{equation}
most of which is concentrated near the inner radius $R$. The radiation
thus encounters total optical depth
\begin{equation} 
\tau_{\rm tot}(R) = \tau(R) + \tau_{\rm sh}(R) \simeq {\kappa
  f_g\sigma^2\over \pi GR}
\end{equation}
whatever the thickness of the shell. Gas distributed in this way is
very optically thick near the black hole when its inner edge $R$ is
small (cf eqn \ref{rtr} below). Then the accretion luminosity $L$ of
the AGN is initially largely trapped and isotropized by scattering,
increasing the interior radiation pressure $P$.
This growing pressure pushes against the weight
\begin{equation}
W(R) = {GM(R)M_g(R)\over R^2} = {4f_g\sigma^4\over G}
\label{weight}
\end{equation}
of the swept--up gas shell at radius $R$ (which is constant with $R$
since $GM(R)M_g(R)/R^2 \propto R.R/R^2$ = constant). 

The Appendix discusses in detail the shell's equation of motion as it
expands. But it is already clear that the effectiveness of radiation
pressure is eventually limited because the shell's optical depth falls
off like $1/R$ as it expands. The force exerted by the radiation drops
as it begins to leak out of the cavity, until for $\tau_{\rm tot}(R)
\sim 1$ it is unable to drive the shell further.

This shows that the sweeping up of gas by radiation pressure must stop
at a `transparency radius'
\begin{equation}
R_{\rm tr} \sim {\kappa f_g\sigma^2\over \pi G} \simeq
50\left({f_g\over 0.16}\right)\sigma_{200}^2~{\rm pc},
\label{rtr}
\end{equation}
where (up to a logarithmic factor) the optical depth $\tau_{\rm tot}$
is of order 1, so that the radiation just escapes, acting as a safety
valve for the otherwise growing radiation pressure. Here $\sigma_{200}
= \sigma/200~{\rm km\, s^{-1}}$.

The Appendix shows that the total gravitational potential energy which the accretion
luminosity must supply to push the galactic gas to this radius is
\begin{equation}
E_{\rm tr} \simeq 3W R_{\rm tr}= {12\kappa f_g^2\sigma^6\over \pi G^2},
\label{energytr}
\end{equation}
so that the central black hole must accrete a mass
\begin{equation}
\Delta M \gtrsim {E_{\rm tr}\over \eta c^2} \sim 3\times10^3\sigma_{200}^6\msun,
\label{deltam}
\end{equation}
where $\eta \simeq 0.1$ is the accretion efficiency.  This is much
smaller than the black hole mass itself, so we expect transparency to
be achieved early in the life of the central SMBH, and easily
maintained after this. In addition, the radiation field of the
accreting SMBH ionizes many of the photoelectric absorbing species
outside this radius, affecting the photoelectric absorption column.

Our discussion so far assumes that the swept--up shell remains
spherical, whereas in reality it is likely to fragment to some
degree. We consider the effects of this further in Section
\ref{discussion} below.

\section{The significance of the transparency radius}

We can rewrite (\ref{rtr}) as
\begin{equation}
R_{\rm tr} = {M_{\sigma}\over M}R_{\rm inf}(M) = R_{\rm inf}(M_{\sigma})
\label{rtr2}
\end{equation}
where 
\begin{equation}
M_{\sigma} = {f_g\kappa\over \pi G^2}\sigma^4
\label{msig}
\end{equation}
is the $M - \sigma$ mass (King, 2003; 2005), and
\begin{equation}
R_{\rm inf}(M) = {GM\over \sigma^2}
\label{rinf}
\end{equation}
is the gravitational influence radius of a hole of mass $M$. From
(\ref{mass}) and the first form of (\ref{rtr}) we also have
\begin{equation}
M(R_{\rm tr}) = 2f_g {f_g\kappa\sigma^4\over \pi G^2} = 2f_gM_{\sigma}
\sim M_{\sigma}
\label{trmass}
\end{equation}
So we can think of the transparency radius $R_{\rm tr}$ as roughly the
radius initially containing a gas mass comparable to the final mass of
the black hole. The pressure of trapped radiation rearranges much of
this gas into a shell at $R_{\rm tr}$. While this is now transparent
to the accretion luminosity of the central SMBH, its enormous mass
makes it a severe obstruction to mechanical outflows. These must shock
against it and effectively stop completely for any SMBH mass below the
critical $M - \sigma$ value. (The significance of the $M_{\sigma}$
mass is that at this point, winds carrying the Eddington thrust of the
SMBH are finally able to drive the gas to large radii, where the wind
shocks no longer cool. The outflow makes a rapid transition to
energy--driving, which largely clears the gas from the galaxy
spheroid, simultaneously halting SMBH growth -- cf King, 2003; 2005.)

The most important outflows are black hole winds driven by radiation
pressure. These have velocities $v \simeq 0.1c$ and momentum scalars
$\dot Mv \simeq L_{\rm Edd}/c$, where $L_{\rm Edd}$ the Eddington
luminosity (King \& Pounds, 2003). The impact of these winds on the
interstellar gas is what ultimately fixes the $M - \sigma$ relation
(King, 2003; 2005). Many of these impacts are likely to occur close at
the transparency radius $R_{\rm tr}$, which we can also write (using
eq. \ref{rtr2}) as
\begin{equation}
R_{\rm tr} = 10^6{M_{\sigma}\over M\sigma_{200}^2}R_s(M)
\label{rtr3}
\end{equation} 
where $R_s = 2GM/c^2$ is the black hole gravitational radius.

\section{Observational constraints}

We argued above that black hole winds are halted by collisions near
$R_{\rm tr}$. The shocked winds must rapidly cool, slow and recombine,
and mix with swept-up ISM. So we expect this gas to have modest
ionization, and much slower velocities than the winds themselves.
These properties are very similar to those inferred for the so--called
Warm Absorber (WA) components in AGN spectra, and we suggest that WA
result from these wind impacts. Here we look for observational tests
of this idea.

Powerful highly ionized winds in AGN have been widely observed in
X--ray spectra over the past decade (Pounds et al., 2003, Reeves et
al., 2003, Tombesi et al., 2010). The radial location $R$ of an 
AGN outflow component is notoriously difficult to determine from the
quantities usually measured -- ionization parameter $\xi$ and
equivalent hydrogen column density $N_H$. These both involve the
electron density, and the column an unknown filling factor as well. To
date this degeneracy has been resolved in very few cases, including
the fast outflow of NGC 4051 (Pounds \& King, 2013).

Tombesi et al. (2013, hereafter T13) provide constraints on the
radial location of warm absorbers, and set this in a wider context.
They consider a sample of 35 type 1 Seyferts previously included in a
study of ultra fast outflow (UFOs), many of which also show
WAs. Although a direct determination is not possible, T13 constrain
the radial locations of all the velocity components in their sample.
They get minimum values from the assumption that the gas is moving at
the local escape velocity, and maximum values from the relation
\begin{equation}
R = {L_i\over N_H\xi}.
\label{ion}
\end{equation}
Here $L_i$ is the ionizing luminosity, and T13 use values of $N_H$ and
$\xi$ from XSTAR spectral fitting.  Figure 1 is based on a figure from
T13 and compares the derived kinetic energy rates with the radial
distance constraints for both UFOs and WAs in the sample. To make the
plot independent of black hole mass the energy rate is in units of
$L_{\rm Edd}$, and the radius in Schwarzschild radii $R_{s}$. As
expected, UFOs cluster at $\sim 10^2R_s$. Importantly, we see that WAs
cluster between $10^{6}-10^{7}$ Schwarzschild radii $R_{s}$.

\begin{figure}
  \centering \includegraphics[width=7cm,angle=0]{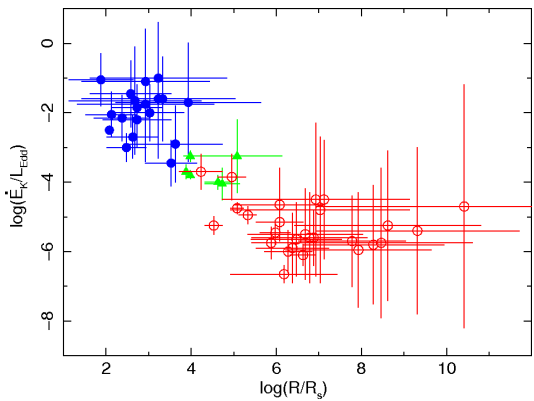}
\caption {Comparison between the radial distance $\log(R/Rs)$ and the
  estimated outflow kinetic energy rate for an overlapping sample of
  WAs (red) and UFOs (blue) in nearby, bright AGN (from Tombesi
  \et\ 2013). For a black hole mass of $10^{8}\msun$ the distance
  scale may be converted to parsecs by noting that $10^6R_s$ is
  $\sim 10$~pc}
\end{figure} 

For comparison, eqn. (\ref{rtr}) predicts $R_{\rm tr}/R_{s} \simeq
5\times 10^{6}$, taking $\sigma =200$~km s$^{-1}$ for an SMBH mass of
$10^{8}\msun$.  Although the spread of data points is broad, with
substantial uncertainty for individual radii, the coincidence of the
radial distance distribution for the full sample of WAs with the
transparency radius derived above is a strong indication of a physical
connection.

A second feature of Fig. 1 is the much higher kinetic power carried by
the UFOs as a group, underlining the importance of high speed winds
for AGN feedback. We note that T13 also show that the measured mass
outflow rates are essentially constant with radius (their Fig. 2). So
the factor $\sim 10^{4}$ between the UFO and WA energy rates in
Fig. 1 is consistent with the idea that these two components characterize
the start and end points of the same mass--conserving outflows, with
mean velocity differences of order $\sim 100$.

UFOs have typical velocities $\sim 0.1c$ characteristic of escape from
$R \sim 100R_s$ (cf King \& Pounds, 2003), and must retain them until
they hit an obstruction with comparable inertia. Since UFOs are also
hypersonic, any deceleration must involve strong shocks. At this point
they are likely to lose almost all of their kinetic energy, since the
shocks are probably strongly Compton--cooled by the AGN radiation
field (King, 2003; 2010). It is natural to assume that the obstruction
is usually the surrounding ISM at $R_{\rm tr}$. In some cases the
winds may hit previous shocked ejecta or infalling gas well within
$R_{\rm tr}$. This latter possibility seems likely for NGC 4051
(Pounds and King 2013 and references therein).

\section{Discussion}
\label{discussion}

We have seen that the trapped radiation pressure exerted by an
accreting supermassive hole is likely to affect the interstellar
medium in its immediate neighbourhood quite strongly. Much of the gas
originally surrounding the hole is swept into a dense shell with
characteristic radius $R_{\rm tr}$ where photons can escape
and prevent a further buildup of radiation pressure inside. This shell
has a mass directly comparable to the final ($M - \sigma$) mass the
hole will eventually reach. 

It appears that the radius $R_{\rm tr}$ is similar to the size
of the region responsible for warm absorber behaviour. This is very
reasonable, since the shell at $R_{\rm tr}$ is so massive that outflows
from the central SMBH must be halted in shocks there if they do not
collide with other structures within $R_{\rm tr}$.

We have so far treated the swept--up shell as spherical and
continuous. It is likely that in practice instabilities can fragment
it before it reaches $R_{\rm tr}$. If the topology of the fragmented
shell remained simple (i.e. a punctured ball) this might relieve the
excess radiation pressure and let the fragmented shell settle at a
radius within $R_{\rm tr}$. However it is likely that the fragmented
shell becomes complex, because is is effectively Rayleigh--Taylor
unstable. The instabilities then produce overturning motions and hence
overlapping gas, which in practice make it difficult for photons to
escape without multiple scattering.  Moreover the undisturbed ISM
immediately outside $R_{\rm tr}$ contributes at least as much opacity
as the swept--up shell. This suggests that even given fragmentation,
$R_{\rm tr}$ is likely to remain a characteristic radius for the
central AGN. An indication of the complex topology of this region
may be that the warm absorption column often has no accompanying
cold absorption, as we might naively expect for a smoothly stratified shell.

In all cases it seems very likely that some of the AGN
luminosity gets absorbed and reradiated by gas with significant
optical depth situated at radii $\lesssim R_{\rm tr}$. If the
reradiated component is roughly blackbody, we find a characteristic
temperature
\begin{equation} 
T_{\rm tr} = \left({lL_{\rm Edd}\over 4\pi fR_{\rm tr}^2\sigma_{\rm
    SB}}\right)^{1/4} \sim 100 \left({lM\over
  fM_{\sigma}}\right)^{1/4}~{\rm K},
\label{ttr} 
\end{equation}
where $l$ is the fraction of the Eddington luminosity $L_{\rm Edd}$
which is reradiated, and $f \times 4\pi $ the solid angle of the
obscuring shell. Given this low temperature and the large photosphere
this component may have evaded detection. A completely intact shell
($f = 1$) might totally obscure an AGN. On either count it seems
possible both that the bolometric luminosities of some known AGN may
have been underestimated, and that some accreting SMBH have escaped
observation entirely.

\section*{Acknowledgments}

We thank Francesco Tombesi for permission to reproduce Figure
1. Astrophysics research at Leicester is supported by an STFC
Consolidated Grant. We thank the referee, Christopher Reynolds,
for a very helpful report.


\section*{Appendix: motion of a gas shell swept 
up by trapped radiation pressure}

Assuming that the swept--up optically thick gas shell is geometrically
thin, its equation of motion is
\begin{equation}
{{\rm d}\over {\rm d}t}[M_g(R)\dot R] = 4\pi R^2P - W
\label{motion}
\end{equation}
Since the radiation pressure does work on the surroundings, we also need
the energy equation
\begin{equation}
{{\rm d}\over {\rm d}t}[VU] = L - 4\pi R^2\dot R P - W\dot R
\label{energy}
\end{equation}
where $V = 4\pi R^3/3$ is the volume interior to the shell, which is filled
with radiation of energy density $U = 3P$, and is supplied with further energy at
the rate $L$. This form is very similar to the energy equation for case of a
wind with mechanical luminosity $L$, assuming that none of this is lost in
cooling after shocking against the surroundings (`energy--driven flow').  The equation
for this case is derived in King (2005) (see also King et al., 2011). We follow the derivation given there, for a general adiabatic relation $P = (\gamma - 1)U$, where the index $\gamma = 4/3, 5/3$ for the present case of radiation and the earlier case of a monatomic gas. We use (\ref{motion}) to eliminate the pressure $P$ from (\ref{energy}). The result is
\begin{eqnarray}
L = {2f_g\sigma^2\over 3G(\gamma -1)}[R^2\dddot R+ (3\gamma + 1)R\dot R\ddot R+(3\gamma -2)\dot R^3] \\ \nonumber
+ {6\gamma - 5\over 3\gamma -3}.{4f_g\sigma^4\over G}\dot R
\label{energy2}
\end{eqnarray}
This reduces to the equation for energy--driving by a wind given in King (2005) and King et al. (2011) if we set $\gamma = 5/3$ (note that the mechanical luminosity $L$ of the wind is $(\eta/2)$ times the near--Eddington radiative luminosity $\sim L_{\rm Edd}$ driving it in this case).

In the trapped radiation case of the present paper, we have $\gamma = 4/3$, giving
\begin{equation}
L = {2f_g\sigma^2\over G}[R^2\dddot R+ 5R\dot R\ddot R+2\dot R^3] +
{12f_g\sigma^4\over G}\dot R.
\label{energy3}
\end{equation}
As in the wind case (see King, 2005; King et al., 2011) there is a constant--velocity solution
$R = v_et$, with
\begin{equation}
L = {4f_g\sigma^2v_e^3\over G} + {12f_g\sigma^4\over G}v_e
\label{ve}
\end{equation}
which is an attractor. This equation defines a unique solution $v_e$.  We can write $L$ as
\begin{equation}
L = {{\rm d}E\over {\rm d} R}v_e
\end{equation}
where $E$ is the total radiation energy inside $R$, so that (\ref{ve}) becomes
\begin{equation}
{{\rm d}E\over {\rm d} R} = \left(3 + {v_e^2\over \sigma^2}\right)W
\label{total}
\end{equation}
 For modest accretion luminosities
$L$ (i.e. well below the Eddington value for the final black holes mass $M_{\sigma}$) we must have $v_e << \sigma$. Then (\ref{total}) implies that the total accretion energy used to
push the gas to the transparency radius $R_{\rm tr}$ is 
\begin{equation}
E_{\rm tr} \simeq 3W R_{\rm tr}= {12\kappa f_g^2\sigma^6\over \pi G^2},
\end{equation}
(cf eqn \ref{energytr}) in the body of the paper.
\end{document}